\documentclass[a4paper,twoside]{article}
\usepackage{amsmath}
\usepackage{amsfonts}
\usepackage{amssymb}
\usepackage[british]{babel}
\usepackage[sans]{dsfont}
\usepackage{cite}
\usepackage{amsthm}

\newcommand{\calH}{\mathcal{H}}
\newcommand{\calL}{\mathcal{L}}
\newcommand{\calT}{\mathcal{T}}
\newcommand{\calI}{\mathcal{I}}

\newcommand{\calF}{\mathcal{F}}

\newcommand{\calM}{\mathcal{M}}
\newcommand{\calS}{\mathcal{S}}
\newcommand{\calU}{\mathcal{U}}

\newcommand{\rmd}{\mathrm{d}}
\newcommand{\rmi}{\mathrm{i}}

\newcommand{\rmq}{\mathrm{q}}
\newcommand{\rmc}{\mathrm{c}}

\newcommand{\Tr}{\operatorname{Tr}}

\newcommand{\E}{\operatorname{\mathbb{E}}}
\newcommand{\Var}{\operatorname{Var}}
\newcommand{\openone}{\mathds1}

\theoremstyle{plain}
\newtheorem{theorem}{Theorem}
\newtheorem{cor}[theorem]{Corollary}
\newtheorem{lemma}[theorem]{Lemma}
\newtheorem{prop}[theorem]{Proposition}
\theoremstyle{definition}

\theoremstyle{remark}
\newtheorem{remark}{Remark}

\begin{document}

\title{Information gain in \\ quantum continual measurements}

\author{Alberto Barchielli
\\
{\small Politecnico di Milano, Dipartimento di Matematica,} \\ {\small and Istituto Nazionale
di Fisica Nucleare, Sezione di Milano.}
\\
{\small E-mail: Alberto.Barchielli@polimi.it}
\\
\and Giancarlo Lupieri
\\
{\small Universit\`a degli Studi di Milano, Dipartimento di Fisica,}\\ {\small and Istituto
Nazionale di Fisica
Nucleare, Sezione di Milano.}
\\ {\small E-mail: Giancarlo.Lupieri@mi.infn.it}}
\date{November 30, 2006}
\maketitle

\begin{abstract}
Inspired by works on information transmission through quantum channels, we propose the use of
a couple of mutual entropies to quantify the efficiency of continual measurement schemes in
extracting information on the measured quantum system. Properties of these measures of
information are studied and bounds on them are derived.
\end{abstract}

\section{Quantum measurements and entropies}\label{sec:intro}
We speak of quantum continual measurements when a quantum system is taken under observation
with continuity in time and the output is not a single random variable, but rather a
stochastic process \cite{BarLP83,BarB91}. The aim of this paper is to quantify, by means of
entropic quantities, the effectiveness of a continual measurement in extracting information
from the underlying quantum system.

Various types of entropies and bounds on informational quantities can be introduced and
studied in connection with continual measurements \cite{Bar01,BarL04,BarLlevico}. In
particular, in Ref.\ \cite{BarLlevico} the point of view was the one of information
transmission: the quantum system is a channel in which some information is encoded at an
initial time; the continual measurement represents the decoding apparatus. In this paper,
instead, we consider the quantum system in itself, not as a transmission channel, and we
propose and study a couple of mutual entropies giving two indexes of how good is the continual
measurement in extracting information about the quantum system.

\subsection{Algebras, states, entropies}
From now on $\calH$ will be a separable complex Hilbert space, the space where our quantum
system lives.

\subsubsection{Von Neumann algebras and normal states}\label{sec:vNa}
A normal state on $\calL(\calH)$ (bounded linear operators on $\calH$) is identified with a
statistical operator, $\calT(\calH)$ and $\calS(\calH)\subset \calT(\calH)$ are the
trace-class and the space of the statistical operators on $\calH$, respectively.

Let $(\Omega,\calF,Q)$ be a measure space, where $Q$ is a $\sigma$-finite measure. We consider
the $W^*$-algebras $L^\infty(\Omega,\calF,Q)$ and
$L^\infty\big(\Omega,\calF,Q;\calL(\calH)\big) \simeq L^\infty(\Omega,\calF,Q)\otimes
\calL(\calH)$. Let us note that a normal state on $L^\infty(\Omega,\calF,Q)$ is a probability
density with respect to $Q$, while a normal state $\sigma$ on
$L^\infty\big(\Omega,\calF,Q;\calL(\calH)\big)$ is a measurable function $\omega \mapsto
\sigma(\omega)\in \calT(\calH)$, $\sigma(\omega)\geq 0$, such that $\Tr\{\sigma(\omega)\}$ is
a probability density with respect to $Q$.

\subsubsection{Relative entropy}
The general definition of the relative entropy $S(\Sigma|\Pi)$ for two states $\Sigma$ and
$\Pi$ is given in [\citen{OhyP93}]; here we give only some particular cases of the general
definition.

Let us consider two quantum states $ \sigma,\,\tau\in \calS(\calH)$ and two classical states
$q_k$ on $L^\infty(\Omega,\calF,Q)$ (two probability densities with respect to $Q$). The von
Neumann entropy, the quantum relative entropy and the classical one are
\begin{gather}\label{relqentropy}
S_\rmq(\tau):= - \Tr \{\tau \ln \tau\}, \qquad S_\rmq(\sigma\|\tau)=\Tr\{\sigma(\ln \sigma-\ln
\tau)\},
\\
S_\rmc(q_1\|q_2)= \int_\Omega Q(\rmd \omega)\,q_1(\omega) \ln\frac{q_1(\omega)}{q_2( \omega)}
\,.
\end{gather}

Let us consider now two normal states $\sigma_k$  on
$L^\infty\big(\Omega,\calF,Q;\calL(\calH)\big)$ and set $q_k(\omega):= \Tr
\{\sigma_k(\omega)\}$, $\varrho_k(\omega):= \sigma_k(\omega)/q_k(\omega)$ (these definitions
hold where the denominators do not vanish and are completed arbitrarily where the denominators
vanish). Then, the relative entropy is
\begin{equation}\label{cqS}
\begin{split}
S(\sigma_1\|\sigma_2) &= \int_\Omega Q(\rmd \omega) \Tr\left\{\sigma_1(\omega)\big(\ln
\sigma_1(\omega)-\ln \sigma_2(\omega)\big)\right\}
\\
{}&= S_\rmc(q_1\|q_2) + \int_\Omega Q(\rmd \omega) \,
q_1(\omega)S_\rmq\big(\varrho_1(\omega)\| \varrho_2(\omega)\big).
\end{split}
\end{equation}

We are using a subscript ``c'' for classical entropies, a subscript ``q'' for purely quantum
ones and no subscript for general entropies, eventually of a mixed character. Having used the
natural logarithm in these definitions, the entropies are in \emph{nats}. To obtain entropies
in \emph{bits} one has to divide by $\ln 2$.

The following result is very useful (\cite{OhyP93} Corollary 5.20 and Eq.\ (5.22)).
\begin{prop}\label{prop:tensorstate}
Let $\Pi_1\otimes \Pi_2$ and $\Sigma_{12}$ be normal states of the tensor product von Neumann
algebra $\mathcal{M}_1\otimes\mathcal{M}_2$ and let
$\Sigma_i=\Sigma_{12}\big|_{\mathcal{M}_i}$, $i=1,2$. Then,
\begin{equation}
\begin{split}
S(\Sigma_{12}\|\Pi_1\otimes \Pi_2)&= S(\Sigma_{1}\|\Pi_1)+S(\Sigma_{12}\|\Sigma_1\otimes
\Pi_2)
\\ {}&=
S(\Sigma_{1}\|\Pi_1)+ S(\Sigma_{2}\|\Pi_2)+S(\Sigma_{12}\|\Sigma_1\otimes \Sigma_2)
\end{split}
\end{equation}
\end{prop}
The quantity $S(\Sigma_{12}\|\Sigma_1\otimes \Sigma_2)$ is the relative entropy of a state
with respect to its marginals; this is what we call \emph{mutual entropy}.

\subsection{Instruments and channels}\label{sec:channels}
\subsubsection{Channels}
Let $\calM_1$ and $\calM_2$ be two $W^*$-algebras. A linear map $\Lambda^*$ from $\calM_2$ to
$\calM_1$ is said to be a \emph{channel} ([\citen{OhyP93}] p.\ 137) if it is completely
positive, unital (i.e.\ identity preserving) and normal (or, equivalently, weakly$^*$
continuous).

Due to the equivalence\cite{Dix57} of w$^*$-continuity and existence of a preadjoint
$\Lambda$, a \emph{channel} is equivalently defined by: $\Lambda$ is a norm-one, completely
positive linear map from the predual $\calM_{1*}$ to the predual $\calM_{2*}$. Let us note
also that $\Lambda$ maps normal states on $\calM_1$ into normal states on $\calM_2$.

A key result which follows from the convexity properties of the relative entropy is
\emph{Uhlmann monotonicity theorem} ([\citen{OhyP93}], Theor.\ 1.5 p.\ 21), which implies that
channels decrease the relative entropy.
\begin{theorem}\label{Uhltheo}
If $\Sigma$ and $\Pi$ are two normal states on $\calM_1$ and $\Lambda$ is a channel from
$\calM_{1*}\to \calM_{2*}$, then $ S(\Sigma\| \Pi)\geq S(\Lambda[\Sigma]\| \Lambda[\Pi])$.
\end{theorem}

\subsubsection{Instruments and POV measures}
The notion of instrument is central in quantum measurement theory; an instrument gives the
probabilities and the state changes \cite{Dav76,Oza84}.

Let $( \Omega,\calF)$ be a measurable space. An  \emph{instrument} $\calI$ is a map valued
measure such that (i) $\calI(F)$ is a completely positive, linear, bounded operator on
$\calT(\calH)$, $\forall F\in \calF$, (ii) $\calI(\Omega)$ is trace preserving, (iii) for
every countable family $\{F_i\}$ of disjoint sets in $\calF$ one has $ \sum_i \Tr\left\{a\,
\calI(F_i)[\rho]\right\}= \Tr\left\{ a\,\calI\left(\bigcup_i F_i\right)[\rho] \right\}$,
$\forall \rho\in \calT(\calH)$, $\forall a\in \calL(\calH)$.

The map $F \mapsto \calI(F)^*[\openone]$ turns out to be a positive operator valued (POV)
measure (the observable associated with the instrument $\calI$). For every $\rho\in
\calS(\calH)$ the map $F \mapsto P_\rho(F):=\Tr \{ \calI(F)[\rho]\}$ is a probability measure:
the probability that the result of the measurement be in $F$ when the pre-measurement state is
$\rho$. Moreover, given the result $F$, the post-measurement state is $\big(
P_\rho(F)\big)^{-1} \calI(F)[\rho]$.

\subsubsection{The instrument as a channel}\label{sec:inst=ch}
Given an instrument $\calI$ with value space $(\Omega,\calF)$ it is always possible to find a
$\sigma$-finite measure on $(\Omega,\calF)$ (or even a probability measure), such that all the
probabilities $P_\rho$, $ \rho \in \calS(\calH)$, are absolutely continuous with respect to
$Q$.

\begin{theorem}[\cite{BarL05QP}, Theorem 2] Let $\calI$ be an instrument on the trace-class of
a complex separable Hilbert space $\calH$ with value space $(\Omega,\calF)$ and let $Q$ be a
$\sigma$-finite measure on $(\Omega,\calF)$ such that $\Tr\{\calI(\bullet)[\rho]\}\ll Q$,
$\forall \rho \in \calS(\calH)$. Then, there exists a unique channel $\Lambda_{\calI}$ from
$\mathcal{T}(\mathcal{H})$ into $L^1\big(\Omega,\calF,Q;\calT(\calH)\big)$ such that
\begin{gather}\label{eq:instr-chann}
\E_Q\big[f\Tr\left\{a\,\Lambda_\calI [\rho]\right\}\big]= \int_\Omega
f(\omega)\Tr\left\{a\,\calI(\rmd \omega)[\rho]\right\} \\  \forall \rho \in \calT(\calH),
\quad \forall a \in \calL(\calH), \quad \forall f\in L^\infty(\Omega,\calF,Q).\notag
\end{gather}
Viceversa, a channel $\Lambda$ from $\mathcal{T}(\mathcal{H})$ into
$L^1\big(\Omega,\calF,Q;\calT(\calH)\big)$ defines a unique instrument $\calI$ by
\begin{equation}
\calI(F)[\rho]=\E_Q\big[1_F\Lambda[\rho]\big], \qquad \forall \rho \in \calT(\calH), \quad
\forall F\in \calF\,.
\end{equation}
\end{theorem}

\subsubsection{A posteriori states}\label{sec:apost}
When $\rho\in \calS(\calH)$, then $\Lambda_\calI[\rho]$ is a normal state on
$L^\infty\big(\Omega,\calF,Q;\calL(\calH)\big)$. Let us normalize the positive trace-class
operators $\Lambda_\calI[\rho](\omega)$ by setting
\begin{equation}\label{apstates}
\pi_\rho(\omega):= \begin{cases} \left(\Tr_{\calH} \left\{\Lambda_\calI[\rho](\omega) \right\}
\right)^{-1} \Lambda_\calI[\rho](\omega) & \text{if }
\Tr_{\calH}\left\{\Lambda_\calI[\rho](\omega) \right\}>0
\\
\tilde \rho \quad \big(\tilde \rho\in \calS(\calH)\text{, fixed} \big) & \text{if }
\Tr_{\calH}\left\{\Lambda_\calI[\rho](\omega) \right\}=0
\end{cases}
\end{equation}
Then, we have
\begin{equation}\label{intapstates}
\int_F \pi_\rho(\omega) P_\rho(\rmd \omega)=\calI(F)[\rho]\,, \quad \forall F\in \calF\,,
\qquad \text{(Bochner integral)}.
\end{equation}
According to Ozawa \cite{Oza85a}, $\pi_\rho$ is a family \emph{of a posteriori states} for the
instrument $\calI$ and the pre-measurement state $\rho$. The interpretation is that
$\pi_\rho(\omega)$ is the state just after the measurement to be attributed to the quantum
system if the result of the measurement has been exactly $\omega$.

Let us note that $p_\rho:=\Tr\left\{\Lambda_\calI[\rho]\right\}$ and $\overline{\pi}_\rho :=
\int_\Omega P_\rho(\rmd \omega) \pi_\rho(\omega)=\mathcal{I}(\Omega)[\rho]$ are the marginals
of the state $ \Lambda_\calI[\rho]$ on the algebras $L^\infty(\Omega,\calF,Q)$ and
$\calL(\calH)$, respectively. Then,
$S\left(\Lambda_\calI[\rho]\|p_\rho\overline\pi_\rho\right)$ is a first example of a mutual
entropy. From Eqs.\ \eqref{cqS} and \eqref{relqentropy} we get
\begin{equation}\label{eq:chi}
S\left(\Lambda_\calI[\rho]\|p_\rho\overline\pi_\rho\right) = \int_\Omega
S_\rmq\big(\pi_\rho(\omega)\| \overline\pi_\rho\big)P_\rho(\rmd \omega) =
S_\rmq(\overline\pi_\rho) - \int_\Omega S_\rmq\big(\pi_\rho(\omega)\big) P_\rho(\rmd \omega).
\end{equation}
Quantities like this one are used in quantum information transmission and are known as  Holevo
capacities or $\chi$-quantities \cite{Hol73,HolS04,BarL06qic}; Eq.\ \eqref{eq:chi} gives the
$\chi$-quantity of the \emph{ensemble} of states $\{P_\rho,\pi_\rho\}$.

\section{Continual measurements} Quantum continual measurement theory can be formulated in
different equivalent ways. To construct our entropic measures of efficiency, we need two
approaches to continual measurements: the one based on positive operator valued measures,
instruments, quantum channels \cite{BarLP83,BarHL93,BarLlevico} and the one based on classical
stochastic differential equations (SDE's), known also as quantum trajectory theory
\cite{BarB91,BarL04,BarH95}.

The SDE approach to continual measurements is based on a couple of sto\-chastic equations, a
linear one for random trace-class operators and a non-linear one for random statistical
operators. The two equations are linked by a change of normalization and a change of
probability measure. Both equations have a Hilbert space formulation, particularly suited for
numerical computations. We shall use a simplified version of SDE's for continual measurements
as presented in \cite{BarB91}.

\subsection{The linear equation} Let $H(t), L_l(t), R_j(t), V_k^r(t), J_k(t)$ be bounded
operators on $\calH$; their time dependence is taken to be continuous from the left and with
limits from the right in the strong topology. The indices $k, l, j$ take a finite number of
values; the index $r$ can take infinitely many values, but in this case the series $\sum_r
V_k^{r}(t)^* V_k^r(t)$ is strongly convergent. Let the operator $H(t)$ be self-adjoint,
$H(t)=H(t)^*$, and let us define ($\forall \rho\in \calT(\calH)$)
\begin{gather}
\mathcal{J}_k(t)[\rho] := \sum_r V_k^r(t)\rho V_k^{r}(t)^*\,, \qquad J_k(t):=
\mathcal{J}_k(t)^*[\openone] =\sum_r V_k^{r}(t)^* V_k^r(t)\,,
\\
\mathcal{L}(t):= \mathcal{L}_0(t)+\mathcal{L}_1(t)+ \mathcal{L}_2(t),
\\
\mathcal{L}_0(t)[\rho] := -\rmi [H(t),\rho] + \sum_l \left( L_l(t)\rho L_l(t)^* - \frac 1 2
\left\{L_l(t)^* L_l(t) , \rho \right\} \right),
\\
\mathcal{L}_1(t)[\rho] :=\sum_j \left( R_j(t)\rho R_j(t)^* - \frac 1 2 \left\{R_j(t)^* R_j(t)
, \rho \right\} \right),
\\
\mathcal{L}_2(t)[\rho] :=\sum_k \left( \mathcal{J}_k(t)[\rho] - \frac 1 2 \left\{J_k(t) , \rho
\right\} \right).
\end{gather}
By $[\ ,\ ]$ we denote the commutator and by $\{\ ,\ \}$ the anticommutator.

Then, we introduce a probability space $(\Omega, \mathcal{F},Q)$ where the Poisson processes
$N_k(t)$, of intensity $\lambda_k$, and the standard (continuous) Wiener processes $W_j(t)$
are defined. All the processes are assumed to be independent from the other ones. We introduce
also the two-times natural filtration of such processes:
\begin{equation}
\mathcal{F}^s_t= \sigma\{W_j(u)-W_j(s),\, N_k(v)-N_k(s),\, u,v\in [s,t],\, j,k=1,\ldots\}.
\end{equation}

Having all these ingredients, we can introduce the linear equation of continual measurement
theory, for the a trace-class valued process $\sigma_t$:
\begin{multline}\label{lineareq}
\rmd \sigma_t= \mathcal{L}(t)[\sigma_{t_-}]\rmd t + \sum_j \left( R_j(t)
\sigma_{t_-} + \sigma_{t_-}R_j(t)^*\right) \rmd W_j(t)
\\ {}+
\sum_k \left(\frac{1}{\lambda_k} \, \mathcal{J}_k(t)[\sigma_{t_-}] -\sigma_{t_-}\right) \left(
\rmd N_k(t) - \lambda_k\rmd t\right).
\end{multline}
The initial condition is taken to be a non-random statistical operator: $\sigma_0\equiv
\sigma_{0_-}\in \mathcal{S(H)}$.

The notation $\sigma_{t_-}$ means that, in case there is a jump in the noise at time $t$, the
value just before the jump  $\sigma_{t_-}$ of $\sigma$ has to be taken. More precisely, if the
augmented natural filtration of the noises is considered, the solution can be taken to be
continuous from the right and with limits from the left and $\sigma_{t_-}$ is just the limit
from the left. We prefer not to add the null sets to the natural filtration and by $\sigma_t $
we mean some $\calF^0_t$-adapted version of the solution.

\paragraph{Properties of the solution.}
Let us consider now, for $0 \leq s\leq t$, the von Neumann algebra
$L^\infty\big(\Omega,\calF^s_t,Q;\calL(\calH)\big) \simeq L^\infty(\Omega,\calF^s_t,Q)\otimes
\calL(\calH)$ (cf.\ Section \ref{sec:vNa}) and let us give a name to the set of normal states
on this algebra:
\begin{equation}
\mathcal{S}^s_t:= \left\{ \tau \in L^1\big(\Omega, \mathcal{F}^s_t,Q;
\mathcal{T}(\mathcal{H})\big) :\tau(\omega) \geq 0 \,, \ \int_\Omega \Tr\{\tau(\omega)\}\,
Q(\rmd \omega)=1\right\}.
\end{equation}

\begin{itemize}
\item
First of all, it is possible to prove that $\sigma_t \in \mathcal{S}_t^0$; we can say that the
solution at time $t$ of Eq.\ \eqref{lineareq} is a kind of quantum/classical state.
\item
The marginals of $\sigma_t$ are (cf.\ Section \ref{sec:apost}):
\begin{itemize}
\item The probability density \ $ p_t:= \Tr\{\sigma_t\}$. The probability measure
$p_t(\omega)Q(\rmd \omega)$ will be the physical probability.
\item
The \emph{a priori state at time $t$} \ $\eta_t:= \E_Q[\sigma_t] \in \mathcal{S(H)}$. This is
the state to be attributed at time $t$ to the system when no selection is done and the result
of the measurement has not been taken into account.
\end{itemize}
\item Moreover, we define the random \emph{a posteriori state at time $t$} \ $\displaystyle
\rho_t=\frac 1 {p_t}\, \sigma_t$. This is the state to be attributed at time $t$ to the system
known the result of the measurement up to $t$.
\end{itemize}
Note that \ $p_0=1$, \ $\rho_0=\sigma_0=\eta_0$, \ $\eta_{t_-}=\eta_t=\eta_{t_+}$.

\subsection{Physical probabilities}

A very important property of Eq.\ \eqref{lineareq} is that $p_t$ is a mean one $Q$-martingale,
which implies that
\begin{equation}\label{eq:physprob}
P_t(\rmd \omega) := p_t(\omega)Q(\rmd \omega)\big|_{\mathcal{F}_t^0}
\end{equation}
is a consistent family of probabilities, i.e., if $0\leq t< T$, $P_T(F)=P_t(F)$, $\forall F\in
\calF^0_t$. These are taken as physical probabilities.

From Eq.\ \eqref{lineareq} we have that $p_t$ satisfies the Dol\'eans equation
\begin{equation}
\rmd p_t = p_{t_-}\biggl\{ \sum_j m_j(t) \, \rmd W_j(t) + \sum_k \left(
\frac{\mu_k(t)}{\lambda_k} - 1 \right) \bigl(\rmd N_k(t) - \lambda_k \, \rmd t\bigr) \biggr\},
\end{equation}
where
\begin{equation}\label{eq:mu,nu}
 m_j(t)= \Tr\left\{ \left( R_j(t) + R_j(t)^*\right) \rho_{t_-}\right\}, \qquad \mu_k(t)=
\Tr\left\{ J_k(t) \rho_{t_-}\right\}.
\end{equation}
The solution of this equation, with $p_0=1$, is
\begin{multline}\label{eq:p_t}
p_t= \exp \biggl\{ \sum_j \biggl[ \int_0^t m_j(s)\,\rmd W_j(s) - \frac 1 2 \int_0^t m_j(s)^2
\,\rmd s \biggr]
\\
{}+ \sum_k \biggl[ \int_0^t \ln \frac{\mu_k(s)}{\lambda_k} \,\rmd N_k(s) +  \int_0^t
\left(\lambda_k - \mu_k(s) \right)\rmd s \biggr]\biggr\}.
\end{multline}

\begin{remark}\label{rem:output}
\begin{enumerate}
\item
The output of the continual measurement is the set of processes $W_j(t)$, $N_k(t)$, $0\leq t
\leq T$, under the physical probability $P_T$; $T$ is a completely arbitrary large time. By
the consistency of the probabilities \eqref{eq:physprob}, $P_T$ can be substituted by $P_t$ in
any expectation involving $\calF^0_t$-measurable random variables (for $t<T$).
\item
By Girsanov theorem and its generalizations for situations with jumps, we have that, under the
physical probability, the processes
\begin{equation}
\widehat W_j(t)= W_j(t) - \int_0^t m_j(s)\, \rmd s
\end{equation}
are independent, standard Wiener processes and $N_k(t)$ is a counting process of stochastic
intensity $\mu_k(t)\rmd t$.
\item
Expressions for the moments of the outputs can be given; in particular we have the mean values
\begin{equation}
\E_{P_t}\left[W_j(t)\right]=\int_0^t n_j(s)\,\rmd s\,, \qquad
\E_{P_t}\left[N_k(t)\right]=\int_0^t \nu_k(s)\,\rmd s\,,
\end{equation}
where
\begin{subequations}\label{eqs:meanval}
\begin{gather}
n_j(t)= \Tr\left\{ \left( R_j(t) + R_j(t)^*\right) \eta_{t}\right\}= \E_{P_t}[m_j(t)],
\\ \label{eq:meanvaljump}
\nu_k(t)= \Tr\left\{ J_k(t) \eta_{t}\right\}=\E_{P_t}[\mu_k(t)].
\end{gather}
\end{subequations}
\end{enumerate}
\end{remark}

\subsection{The non-linear SDE}
Under the physical law $P_T$, the a posteriori states
$\rho_t$ satisfy the non-linear SDE
\begin{multline}\label{nonlineareq}
\rmd \rho_t= \mathcal{L}(t)[\rho_{t_-}]\rmd t + \sum_j \left( R_j(t) \rho_{t_-} +
\rho_{t_-}R_j(t)^* - m_j(t) \rho_{t_-}\right) \rmd \widehat W_j(t)
\\ {}+
\sum_k \left(\frac{1}{\mu_k(t)} \, \mathcal{J}_k(t)[\rho_{t_-}] -\rho_{t_-}\right) \left( \rmd
N_k(t) - \mu_k(t)\rmd t\right).
\end{multline}

Let us stress that for the a priori states we have
\begin{equation}
\eta_t=\E_Q[\sigma_t]=\E_{P_t}[\rho_t]
\end{equation}
and that they satisfy the \emph{master equation}
\begin{equation}
\frac{\rmd \ }{\rmd t}\,\eta_t=\calL(t)[\eta_t].
\end{equation}

\subsection{The fundamental matrix and the instruments}
To apply the notions of Section \ref{sec:intro} to continual measurements, we need to see how
such a theory is connected to instruments and channels \cite{BarB91,Bar01,BarL04,BarLlevico}.
This is done by introducing the fundamental matrix $\Lambda^s_t$ of \eqref{lineareq}. This
operator is defined by stipulating that $\Lambda^s_t[|u_i\rangle \langle u_j |]$ satisfies
\eqref{lineareq} with initial condition $\Lambda^s_s[|u_i\rangle \langle u_j |]=|u_i\rangle
\langle u_j |$, where $\{u_i\,\,i=1,\ldots\}$ is a c.o.n.s.\ in $\calH$. It turns out that
${\Lambda^s_t}$ is a channel from $\calT(\calH)$ into $L^1\big(\Omega,\calF^s_t,Q;
\calT(\calH)\big)$, or, by trivial ampliation, from $L^1\big(\Omega,\calF^r_s,Q;
\calT(\calH)\big)$ into $L^1\big(\Omega,\calF^r_t,Q; \calT(\calH)\big)$, $0\leq r\leq s\leq
t$. Then, we have
\begin{equation}\label{eq:lambdacomp}
\Lambda^s_t[\sigma_s]=\sigma_t \,, \qquad \Lambda^s_t= \Lambda^u_t\circ \Lambda^s_u \,,\quad
0\leq s \leq u \leq t\,.
\end{equation}
The instrument associated to this channel is
\begin{equation}
\mathcal{I}^s_t(F)[\rho]= \E_Q\left[1_F \Lambda^s_t[\rho]\right] \equiv \int_F
\Lambda^s_t(\omega)[\rho] Q(\rmd \omega), \qquad \forall F\in \mathcal{F}^s_t.
\end{equation}
The time evolution of the quantum states is the one generated by $\calL(t)$ and we have
\begin{equation}
\mathcal{U}(t,s)[\rho]=\mathcal{I}^s_t(\Omega)[\rho]= \E_Q\left[\Lambda^s_t[\rho]\right],
\end{equation}
\begin{equation}
\calU(t,s)[\eta_s]=\eta_t\,, \qquad \calU(t,s)= \calU(t,u)\circ \calU(u,s)\,, \quad 0\leq s
\leq u \leq t\,.
\end{equation}

According to the definitions of Section \ref{sec:apost}, the random statistical operator
$\rho_t$ is the a posteriori state for the instrument $\mathcal{I}^0_t$ and the
pre-measurement state $\rho_0\equiv \eta_0$.

Another important property is
\begin{equation}\label{eq:meanonthepast}
\E_Q[\sigma_t|\calF^s_t] =\Lambda^s_t[\eta_s]\in \calS^s_t\,.
\end{equation}
Indeed, by the first of \eqref{eq:lambdacomp} and the fact that $\Lambda^s_t$ is
$\calF^s_t$-measurable, we have $ \E_Q[\sigma_t|\calF^s_t]
=\Lambda^s_t\big[\E_Q[\sigma_s|\calF^s_t] \big]$. By the fact that all the noises have
independent increments, we have that $\sigma_s$ is independent from $\calF^s_t$ and
$\E_Q[\sigma_s|\calF^s_t]=\E_Q[\sigma_s]=\eta_s$. This gives Eq.\ \eqref{eq:meanonthepast}.

\section{Mutual entropies and information gains}
\subsection{The information embedded in the a posteriori states}\label{sec:Sapost}
The quantity $\sigma_t$ is a state on $L^\infty\big(\Omega,\calF^0_t,Q; \calL(\calH)\big)=
L^\infty(\Omega,\calF^0_t,Q)\otimes \calL(\calH)$ and its marginals on
$L^\infty(\Omega,\calF^0_t,Q)$ and $\calL(\calH)$ are $p_t$ and $\eta_t$, respectively. The
mutual entropy $S(\sigma_t\|p_t\eta_t)$ is the ``information'' contained in the joint state
with respect to the product of these marginals; more explicitly we have (compare with
\eqref{eq:chi})
\[
S(\sigma_t\|p_t\eta_t)=\int_\Omega P_t(\rmd \omega) \Tr\left\{ \rho_t(\omega) \big(\ln
\rho_t(\omega) - \ln \eta_t\big)\right\}
\]
and we can write
\begin{equation}\label{eq:Sapost}
S(\sigma_t\|p_t\eta_t)=\E_{P_t}[S_\rmq(\rho_t\|\eta_t)]=S_\rmq(\eta_t)-
\E_{P_t}[S_\rmq(\rho_t)].
\end{equation}

This mutual entropy is a sort of quantum information embedded by the measurement in the a
posteriori states. When the measurement is not informative, we have $\rho_t(\omega)=\eta_t$
and $S(\sigma_t\|p_t\eta_t)=0$. It is zero also if for any reason it happens that $\eta_t$ is
a pure state. For instance, if $\calU(t,0)$ has a unique equilibrium state which is pure, then
$\lim_{t\to +\infty} S(\sigma_t\|p_t\eta_t)=0$ even if the measurement is ``good''.

Let us note that from Eq.\ \eqref{eq:Sapost} we have the bound
\begin{equation}
S(\sigma_t\|p_t\eta_t)\leq S_\rmq(\eta_t).
\end{equation}
When the von Neumann entropy of the a priori state is not zero, an instantaneous index of
``goodness'' of the measurement could be $S(\sigma_t\|p_t\eta_t)\big/S_\rmq(\eta_t)$, while a
``cumulative'' index could be $\int_0^T \frac{S(\sigma_t\|p_t\eta_t)} {S_\rmq(\eta_t)}\,\rmd
t$ .

\subsection{A classical continual information gain}
\subsubsection{Product densities}
Let us consider any time $s$ in the time interval $(0,t)$ and let us decompose the von Neumann
algebra $L^\infty(\Omega,\calF^0_t,Q)$ as
$L^\infty(\Omega,\calF^0_t,Q)=L^\infty(\Omega,\calF^0_s,Q)\otimes
L^\infty(\Omega,\calF^s_t,Q)$. Now, the density $p_t$ can be seen as a state on
$L^\infty(\Omega,\calF^0_t,Q)$ and we can consider its marginals $p^0_s$ and $p^s_t$ on the
two factors $L^\infty(\Omega,\calF^0_s,Q)$ and $L^\infty(\Omega,\calF^s_t,Q)$, respectively.
These marginals are given by
\begin{equation}
p^0_s= \E_Q[p_t|\calF^0_s], \qquad p^s_t= \E_Q[p_t|\calF^s_t].
\end{equation}
By using the fact that $\{p_t,\, t\geq0\}$ is a martingale and by taking the trace of Eq.\
\eqref{eq:meanonthepast}, we get
\begin{equation}
p^0_s=p_s\,, \qquad p^s_t=\Tr\{ \Lambda^s_t[\eta_s]\}\,.
\end{equation}
By comparing the last equality with $p_t=\Tr\{\sigma_t\}=\Tr\{ \Lambda^0_t[\eta_0]\}$, we see
that $p^s_t$ is similar to $p_t$, but with $s$ as initial time, instead of $0$, and with
$\eta_s$ as initial state, instead of $\eta_0$. By this remark and Eq.\ \eqref{eq:p_t}, we get
\begin{multline}\label{eq:deltap}
p^s_t= \exp \biggl\{ \sum_j \biggl[ \int_s^t m_j(u;s)\,\rmd W_j(u) - \frac 1 2 \int_s^t
m_j(u;s)^2 \,\rmd u \biggr]
\\
{}+ \sum_k \biggl[ \int_s^t \ln \frac{\mu_k(u;s)}{\lambda_k} \,\rmd N_k(u) + \int_s^t
\left(\lambda_k - \mu_k(u;s) \right)\rmd u \biggr]\biggr\},
\end{multline}
where
\begin{gather}
m_j(t;s)= \Tr\left\{ \left( R_j(t) + R_j(t)^*\right) \rho^s_{t_-}\right\} ,\qquad \mu_k(t;s)=
\Tr\left\{ J_k(t) \rho^s_{t_-}\right\},
\\ \label{eq:deltarho}
\rho^s_t=\frac{1}{p^s_t}\,\Lambda^s_t[\eta_s].
\end{gather}
The random state $\rho^s_t$ is the a posteriori state for the instrument $\calI^s_t$ and the
pre-measurement state $\eta_s$; it satisfy the non-linear SDE \eqref{nonlineareq}.

Then, we can consider the mutual entropy $S_\rmc(p_t\| p^0_sp^s_t)$. But the significance of
this quantity is dubious, because the time $s$ is completely arbitrary and, moreover, we could
divide the time interval in more pieces. For instance, we can take the decomposition
$L^\infty(\Omega,\calF^0_t,Q)=L^\infty(\Omega,\calF^0_r,Q)\otimes L^\infty(\Omega,\calF^r_s,Q)
\otimes L^\infty(\Omega,\calF^s_t,Q)$ and we recognize that $p^0_rp^r_sp^s_t$ is the product
of the marginals of $p_t$ related to this decomposition. Taking a finer generic partition of
$(0,t)$ with $t_0=0$ and $t_n=t$, we recognize that $\prod_{j=1}^n p^{t_{j-1}}_{t_j}$ is again
a product of marginals of $p_t$. To eliminate arbitrariness, let us consider finer and finer
partitions and let us go to a continuous product of marginals.

Let us note that we have
\begin{equation*}
\lim_{s\uparrow t} m_j(t;s)= n_j(t),\qquad \lim_{s\uparrow t}\mu_k(t;s)= \nu_k(t), \qquad
\text{a.s.}
\end{equation*}
Then, for an infinitesimal interval we get
\begin{multline}\label{eq:pinfinitesimal}
p^{s}_{s+\rmd s} = \exp\biggl\{ \sum_j \left[n_j(s) \, \rmd W_j(s) - \frac 1 2 \, n_j(s)^2\rmd
s \right] \\ {}+ \sum_k \left[\frac{\nu_k(s)}{\lambda_k} \rmd N_k(s) +\left( \lambda_k
-\nu_k(s)\right) \rmd s\right] \biggr\}
\end{multline}
and, so, the following density $q_t$ is the continuous product of marginals of $p_t$:
\begin{multline}\label{eq:q_t}
q_t=  \exp \biggl\{ \sum_j \biggl[ \int_0^t n_j(s)\,\rmd W_j(s) - \frac 1 2 \int_0^t
n_j(s)^2 \,\rmd s \biggr]
\\
{}+ \sum_k \biggl[ \int_0^t \ln \frac{\nu_k(s)}{\lambda_k} \,\rmd N_k(s) +  \int_0^t
\left(\lambda_k- \nu_k(s) \right)\rmd s \biggr]\biggr\}.
\end{multline}

Notice that $n_j(t)$ and $\nu_k(t)$ are deterministic functions. Under the probability
$q_T(\omega)Q(\rmd \omega)$, the processes $ W_j(t) - \int_0^t n_j(s)\, \rmd s $ are
independent, standard Wiener processes and $N_k(t)$ is a Poisson process of time dependent
intensity $\nu_k(t)$.

Under $q_T(\omega)Q(\rmd \omega)$, the processes $W_j$, $N_k$ have independent increments as
under $Q$ (so they can be interpreted as noises), but the means have been changed and made
equal to the means they have under $P_T$.

The fact that it is possible to consider a ``continuous product of marginals''  is not so
unexpected; indeed, the theory of continual measurements is connected to infinite divisibility
\cite{BarHL93}.

We have already seen that the marginals of $p_t$ with respect to the decomposition of the time
interval $(0,t)$ into $(0,s)$ and $(s,t)$ are $p^0_s=p_s$ and $p^s_t$ given by Eq.\
\eqref{eq:deltap}. The analogous marginals for $q_t$ are $q^0_s=q_s$ and
\begin{multline}\label{eq:deltaq}
q^s_t=  \exp \biggl\{ \sum_j \biggl[ \int_t^u n_j(s)\,\rmd W_j(s) - \frac 1 2 \int_t^u
n_j(s)^2 \,\rmd s \biggr]
\\
{}+ \sum_k \biggl[ \int_t^u \ln \frac{\nu_k(s)}{\lambda_k} \,\rmd N_k(s) +  \int_t^u
\left(\lambda_k- \nu_k(s) \right)\rmd s \biggr]\biggr\}=\frac {q_t}{q_s}\,.
\end{multline}

\subsubsection{The classical mutual entropy $S_\rmc(p_t\|q_t)$}\label{sec:Scontinual}
The density $q_t$ is no more dependent on some arbitrary choice of intermediate times and the
measure $q_T(\omega)Q(\rmd \omega)$ has a distinguished role and can be considered as a
reference measure. So, we can introduce the relative entropy
\[
S_\rmc(p_t\|q_t)= \E_{P_t}\left[\ln \frac{ p_t}{ q_t}\right].
\]
Being $q_t$ a product of marginals of $p_t$, this quantity is a mutual entropy and, being
$q_t$ the finest product of marginals, we can interprete $S_\rmc(p_t\|q_t)$ as a measure of
the classical information on the measured system extracted in the time interval $(0,t)$. Other
reasons can be given to reinforce this interpretation.

By Eqs.\ \eqref{eq:p_t}, \eqref{eq:deltap}, \eqref{eq:q_t}, \eqref{eq:deltaq} we have
$p^0_t=p_t$, $q^0_t=q_t$, $q_u=q_tq^t_u$. By Proposition \ref{prop:tensorstate} or by direct
computation, we get
\begin{equation}\label{eq:incrementofinfo}
S_\rmc(p_t\|q_t)-S_\rmc(p_s\|q_s) =S_\rmc(p_t\|p_sq^s_t),\qquad 0\leq s\leq t\,.
\end{equation}
Firstly, by the positivity of relative entropies, this equation says that
\begin{equation}\label{eq:increasing}
0\leq S_\rmc(p_s\|q_s)\leq S_\rmc(p_t\|q_t),
\end{equation}
i.e.\ that $S_\rmc(p_t\|q_t)$ is non negative and not decreasing in time, as should be for a
measure of an information gain in time. Moreover, the increment of information in the time
interval $(s,t)$ can be written as
\begin{equation}
S_\rmc(p_t\|p_sq^s_t)=\E_Q \left[ p_s \E_Q\left[ \frac{p_t}{p_s} \ln \frac{p_t/p_s} {q_t/q_s}
\bigg|\calF^0_s\right]\right] .
\end{equation}
This expression can be interpreted as a \emph{conditional relative entropy} (\cite{CovT91}
pp.\ 22--23). The quantity $\E_Q\left[ \frac{p_t}{p_s} \ln \frac{p_t/p_s} {q_t/q_s}
\Big|\calF^0_s\right]$ has the same structure as $S_\rmc(p_t\|q_t)$, but it refers to the
interval $(s,t)$ and it is constructed with the conditional densities. We can say that Eq.\
\eqref{eq:incrementofinfo} expresses in a consistent way a kind of ``additivity property'' of
our measure of information.

Having the explicit exponential forms of the densities $p_t$ and $q_t$, we can compute the
explicit expression of the information gain.
\begin{prop}
The explicit expression of the classical mutual entropy $S_\rmc(p_t\|q_t)$ is
\begin{equation}\label{eq:explS_c}
S_\rmc(p_t\|q_t)=\frac 1 2 \sum_j \int_0^t \Var_{P_t}[m_j(s)]\rmd s + \sum_k \int_0^t
\E_{P_t}\left[\mu_k(s) \ln \frac{\mu_k(s)}{\nu_k(s)}\right]\rmd s
\end{equation}
\end{prop}
\begin{proof} By Eqs.\ \eqref{eq:p_t} and \eqref{eq:q_t} we get
\begin{multline*}
\ln \frac{ p_t}{ q_t}= \sum_j \biggl[ \int_0^t \bigl( m_j(s) - n_j(s)\bigr)\rmd W_j(s) - \frac
1 2 \int_0^t \bigl( m_j(s)^2 -n_j(s)^2 \bigr)\rmd s \biggr]
\\
{}+ \sum_k \biggl[ \int_0^t \ln \frac{\mu_k(s)}{\nu_k(s)} \bigl(\rmd N_k(s) - \lambda_k\, \rmd
s \bigr) +  \int_0^t  \left(\lambda_k\ln \frac{\mu_k(s)}{\nu_k(s)} - \mu_k(s)
+\nu_k(s)\right)\rmd s \biggr]
\\ {}
= \sum_j \biggl[ \int_0^t \bigl( m_j(s) - n_j(s)\bigr)\bigl(\rmd W_j(s)- m_j(s)\rmd s \bigr) +
\frac 1 2 \int_0^t \bigl( m_j(s) -n_j(s) \bigr)^2\,\rmd s \biggr]
\\
{}+ \sum_k \biggl[ \int_0^t \ln \frac{\mu_k(s)}{\nu_k(s)} \bigl(\rmd N_k(s) - \mu_k(s)\, \rmd
s \bigr) \\ {} +  \int_0^t  \mu_k(s)\left(\frac {\nu_k(s)} {\mu_k(s)} - \ln
\frac{\nu_k(s)}{\mu_k(s)} - 1\right)\rmd s \biggr].
\end{multline*}
By point 2 in Remark \ref{rem:output}, the first term in the $j$ sum and the first term in the
$k$ sum have zero mean under $P_T$ (or under $P_t$, by consistency). Therefore, Eq.\
\eqref{eq:explS_c} follows by taking the $P_t$-mean of $\ln p_t/q_t$ and by taking into
account Eqs.\ \eqref{eqs:meanval}.
\end{proof}

\begin{remark}\label{rem:S_c}
\begin{enumerate}
\item By \eqref{eq:meanvaljump} and Jensen inequality applied to the convex function
$x\ln x$, we have that both integrands in formula \eqref{eq:explS_c} are non-negative and, so,
we have
\begin{equation}\label{eq:res3}
\frac{\rmd \ }{\rmd t}\, S_\rmc(p_t\|q_t)=\frac 1 2 \sum_j  \Var_{P_t}[m_j(t)]+ \sum_k
\E_{P_t}\left[\mu_k(t) \ln \frac{\mu_k(t)}{\nu_k(t)}\right]\geq 0.
\end{equation}
The positivity of this time derivative follows also from Eq.\ \eqref{eq:increasing}.

\item By the properties of relative entropy $\E_{P_T}[S_\rmq(\rho_t\|\eta_t)]=0$
is equivalent to $\rho_t=\eta_t$, $P_T$-a.s. By Eqs.\ \eqref{eqs:meanval}, \eqref{eq:res3},
this last relation implies the vanishing of the quantity \eqref{eq:res3}. So, we have
\begin{equation}
\E_{P_T}[S_\rmq(\rho_t\|\eta_t)]=0  \quad \Rightarrow \quad \frac{\rmd \ }{\rmd t}\,
S_\rmc(p_t\|q_t)=0\,.
\end{equation}
\item From Eqs.\ \eqref{eq:mu,nu}, \eqref{eqs:meanval}, \eqref{eq:res3} we see that
\begin{itemize}
\item
if $R_j(t)+R_j(t)^*\propto \openone$, then $\Var_{P_t}[m_j(t)]=0$,
\item
if $J_k(t)\propto \openone$, then $\ln \frac{\mu_k(t)}{\nu_k(t)}=0$.
\end{itemize}
This says that when both conditions hold for all $j$ and $k$, no information is extracted from
the system, whatever the initial state is.
\end{enumerate}
\end{remark}

\subsection{A quantum/classical mutual entropy}

The two mutual entropies introduced in Sections \ref{sec:Sapost} and \ref{sec:Scontinual} can
be obtained from a unique mutual entropy
\begin{equation}
S(\sigma_t\|q_t\eta_t)=\int_\Omega Q(\rmd \omega) \Tr\left\{ \sigma_t(\omega) \big(\ln
\sigma_t(\omega) - \ln q_t(\omega)\eta_t\big)\right\}.
\end{equation}
Indeed, by Proposition \ref{prop:tensorstate} or by direct computation, we get
\begin{equation}
S(\sigma_t\|q_t\eta_t)=S(\sigma_t\|p_t\eta_t)+ S_\rmc(p_t\|q_t)=
\E_{P_t}[S_\rmq(\rho_t\|\eta_t)]+ S_\rmc(p_t\|q_t).
\end{equation}

\section{An upper bound on the increments of $S_\rmc(p_t\|q_t)$}
\subsection{The main bound}

By Proposition \ref{prop:tensorstate} and Eqs.\ \eqref{eq:p_t}, \eqref{eq:deltap},
\eqref{eq:q_t}, \eqref{eq:deltaq}, the increment of information in the time interval $(t,u)$
can be expressed as
\begin{equation}\label{eq:useful}
S_\rmc(p_u\|q_u)- S_\rmc(p_t\|q_t)=S_\rmc(p_u\|p_tp^t_u)+S_\rmc(p^t_u\|q^t_u).
\end{equation}

\begin{lemma} \label{lemma:1}
For $0\leq t\leq u$,  we have the bound
\begin{equation}\label{eq:ineq}
0\leq S_\rmc\left(p_u\|p_tp^t_u\right) \leq \E_{P_u}\left[ S_\rmq\left(\rho_t\|\eta_t\right)-
S_\rmq\left(\rho_u\|\rho^t_u\right)\right].
\end{equation}
\end{lemma}
\begin{proof}
Consider the mutual entropy $S(\sigma_t\|p_t\eta_t)$ introduced in Section \ref{sec:Sapost}
and apply to both states the channel $\Lambda^t_u$. By Theorem \ref{Uhltheo} and the
definition \eqref{eq:deltarho} we get the inequality
\begin{multline*}
\E_{P_t}\left[S_\rmq\left(\rho_t\|\eta_t\right)\right]= S\left(\sigma_t\|p_t\eta_t\right) \geq
S\left(\Lambda^t_u[\sigma_t]\|\Lambda^t_u[p_t\eta_t]\right) =
S\left(\sigma_u\|p_t\sigma^t_u\right) \\ {}= S\left(p_u\rho_u\|p_tp^t_u\rho^t_u\right) =
\E_{P_u}\left[ \Tr \left\{ \rho_u \left( \ln p_u + \ln \rho_u -\ln \left( p_tp^t_u\right) -
\ln \rho^t_u \right)\right\}\right]
\\ {}=
S_\rmc\left(p_u\|p_tp^t_u\right) + \E_{P_u}\left[ S_\rmq\left(\rho_u\|\rho^t_u\right)\right],
\end{multline*}
and this gives \eqref{eq:ineq}.
\end{proof}
Apart from the different notations, Eq.\ \eqref{eq:ineq} is the bound (29) in Ref.\
\cite{BarLlevico}.

From Eqs.\ \eqref{eq:pinfinitesimal} and \eqref{eq:deltaq} we get immediately
\begin{equation}
\lim_{u\downarrow t} \frac{S_\rmc\left(p^t_u\|q^t_u\right)} {u-t} = 0\,.
\end{equation}
Then, the second summand in the expression \eqref{eq:useful} of the increment of information
becomes negligible with respect to the first when $u\downarrow t$. Therefore, from Lemma
\ref{lemma:1} we have immediately the following theorem.
\begin{theorem}[The bound on the derivative of $S_\rmc(p_t\|q_t)$]
The following bound holds:
\begin{multline}\label{eq:main}
0\leq \frac{\rmd\ }{\rmd t}\, S_\rmc(p_t\| q_t) \leq - \frac{\rmd\ }{\rmd u}\,
\E_{P_T}[S_\rmq(\rho_u\|\rho^t_u)]\Big|_{u=t^+}
\\ {} \equiv
\frac{\rmd\ }{\rmd t}\,\E_{P_T}[S_\rmq(\rho_t)]- \frac{\rmd\ }{\rmd
u}\,\E_{P_T}[S_\rmq(\rho^t_u)]\Big|_{u=t^+}\,.
\end{multline}
\end{theorem}

\begin{remark} We already saw in Remark \ref{rem:S_c} that
$\E_{P_T}[S_\rmq(\rho_t\|\eta_t)]=0$ is equivalent to $\rho_t=\eta_t$, $P_T$-a.s.; but this
implies $\rho_u= \rho_u^t$, $P_T$-a.s., because in this case these two quantities, which
satisfy the same equation, have the same initial condition at time $t$. Therefore we have
$\E_{P_T}[S_\rmq(\rho_u\|\rho^t_u)]=0$, $\forall u\geq t$, and
\begin{equation}
\E_{P_T}[S_\rmq(\rho_t\|\eta_t)]=0  \quad \Rightarrow \quad  - \frac{\rmd\ }{\rmd u}\,
\E_{P_T}[S_\rmq(\rho_u\|\rho^t_u)]\Big|_{u=t^+}=0.
\end{equation}
\end{remark}

\subsection{Explicit computation of the bound}

All the derivatives can be elaborated and from Eq.\ \eqref{eq:main} we get the
following explicit form of the difference between the bound and the time derivative
in which we are interested in.
\begin{prop} By computation of all the terms appearing in Eq.\ \eqref{eq:main} we get
\begin{multline}\label{eq:explicitb}
0\leq  \frac{\rmd\ }{\rmd t}\,\E_{P_T}[S_\rmq(\rho_t)]- \frac{\rmd\ }{\rmd
u}\,\E_{P_T}[S_\rmq(\rho^t_u)]\Big|_{u=t^+}-\frac{\rmd\ }{\rmd t}\, S_\rmc(p_t\| q_t)
\\ {} =
\sum_k\E_{P_T}\big[\Tr \big\{J_k(t)\rho_t\left(\ln \rho_t- \ln \eta_t\right)
-\mathcal{J}_k(t)[\rho_t]\left(\ln \mathcal{J}_k(t)[\rho_t] -\ln
\mathcal{J}_k(t)[\eta_t]\right) \big\}\big]
\\ {}+ \sum_j\E_{P_T}\left[\Tr \left\{ R_j(t)\eta_t \big[R_j(t)^*,\,\ln \eta_t\big]
-R_j(t)\rho_t \big[R_j(t)^*,\,\ln \rho_t\big]\right\}\right] \\ {}+ \sum_l\E_{P_T}\left[\Tr
\left\{L_l(t)\eta_t\big[ L_l(t)^*,\, \ln \eta_t\big] -L_l(t)\rho_t\big[ L_l(t)^*,\, \ln
\rho_t\big]\right\}\right]
\\ {}+
\frac 1 2 \sum_j \E_{P_T}\bigg[ \int_0^{+\infty}\rmd u\Tr \bigg\{R_j(t)^*\frac
{\eta_t}{u+\eta_t}R_j(t)^*\frac {\eta_t}{u+\eta_t} \\ {} -R_j(t)^*\frac
{\rho_t}{u+\rho_t}R_j(t)^*\frac {\rho_t}{u+\rho_t} \\ {}+ \frac
{\eta_t}{u+\eta_t}R_j(t)\frac {\eta_t}{u+\eta_t}R_j(t) - \frac
{\rho_t}{u+\rho_t}R_j(t)\frac {\rho_t}{u+\rho_t}R_j(t)
\\ {}+\frac {2}{u+\eta_t}R_j(t)\frac {{\eta_t}^2}{u+\eta_t}R_j(t)^* - \frac
{2}{u+\rho_t}R_j(t)\frac {{\rho_t}^2}{u+\rho_t}R_j(t)^* \bigg\}\bigg].
\end{multline}
\end{prop}

\begin{proof} Let us start with the term $\frac{\rmd\ }{\rmd
u}\,\E_{P_T}[S_\rmq(\rho^t_u)]\Big|_{u=t^+}$. By recalling that $\rho^t_u$ satisfies in $u$
the non-linear SDE with initial condition $\eta_t$ at $u=t$ and that $\eta_t+
\calL(t)[\eta_t]\rmd t=\eta_{t+\rmd t}$, we get
\[
\rho^t_{t+\rmd t}-\eta_{t+\rmd t}=  \sum_j A_j(t)\rmd  \check{W}_j(t) + \sum_k \left(
\tau_k(t)-\eta_{t}\right) \left( \rmd N_k(t) - \nu_k(t)\rmd t\right),
\]
where
\[
A_j(t):=R_j(t) \eta_{t} + \eta_{t}R_j(t)^* - n_j(t) \eta_{t}\,, \qquad
\tau_k(t):=\frac{1}{\nu_k(t)} \, \mathcal{J}_k(t)[\eta_{t}],
\]
\[
\rmd \check{W}_j(t) := \rmd  W_j(t)-n_j(t)\rmd t\,.
\]
By setting also
\[
B(t) :=- \frac 1 2 \sum_k\left\{J_k(t)-\nu_k(t),\eta_t \right\}+ \mathcal{L}_0(t) [\eta_t]+
\mathcal{L}_1(t) [\eta_t],
\]
we can write
\[
\eta_{t+\rmd t}= \left(1-\sum_k \nu_k(t)\rmd t \right)\eta_t +\sum_k \nu_k(t) \tau_k(t) \rmd t
+B(t)\rmd t\,.
\]
Moreover, by the properties of the increments of the counting processes, we have
\[
\rho^t_{t+\rmd t}\rmd N_k(t)=\tau_k(t)\rmd N_k(t),
\]
\begin{multline*}
\left(1-\sum_k\rmd N_k(t)\right)\rho^t_{t+\rmd t} \\ {}=\left(1-\sum_k\rmd
N_k(t)\right)\left(\eta_{t}+B(t) \rmd t +\sum_jA_j(t)\,\rmd \check{W}_j(t) \right).
\end{multline*}

By putting these things all together and by using the rules of stochastic calculus, we get
\begin{multline*}
\rho^t_{t+\rmd t}\ln \rho^t_{t+\rmd t}-\eta_t\ln \eta_t= \sum_k \bigl[\tau_k(t)\ln
\tau_k(t)-\eta_t\ln \eta_t\bigr]\rmd N_k(t)
\\ + \eta_t
\left[\ln \left(\eta_{t} +B(t)\rmd t +\sum_jA_j(t)\,\rmd \check{W}_j(t) \right)-\ln
\eta_{t}\right]+\sum_jA_j(t) \ln \eta_{t}\,\rmd \check{W}_j(t)
\\ {}+ B(t) \rmd t \, \ln
\eta_{t} +\sum_jA_j(t) \left[\ln \left(\eta_{t}+A_j(t)\,\rmd \check{W}_j(t) \right)-\ln
\eta_{t}\right]\rmd \check{W}_j(t).
\end{multline*}

It exists a nearly obvious and very useful integral representation of the logarithm of an
operator (\cite{OhyP93} p.\ 51):
\[
\ln A = \int_0^{+\infty} \left( \frac 1 {1+t} - \frac 1 {t+A}\right) \rmd t.
\]
By iterating this formula we get also
\begin{multline*}
\ln(A+B)-\ln A = \int_0^{+\infty}  \frac 1 {t+A} \,B\, \frac 1 {t+A+B}\, \rmd t \\ {}=
\int_0^{+\infty}  \frac 1 {t+A} \,B\,\frac 1 {t+A}\left(1 -B\, \frac 1 {t+A+B}\right) \rmd
t\,.
\end{multline*}
These two formulae and stochastic calculus rules allow to write
\begin{multline*}
\E_{P_{T}}\left[\Tr\left\{\rho^t_{t+\rmd t}\ln \rho^t_{t+\rmd t}-\eta_t\ln
\eta_t\right\}\right]= \sum_k \left[S_\rmq(\eta_t)-
S_\rmq\big(\tau_k(t)\big)\right]\nu_k(t)\rmd t
\\ {}- \rmd t
\sum_j \int_0^{+\infty}\rmd u \Tr\left\{\frac {\eta_t}{(u+\eta_t)^2} \, A_j(t)\, \frac
{1}{u+\eta_t}\, A_j(t)\right\}
\\ {}+
\rmd t \sum_j \int_0^{+\infty}\rmd u \Tr\left\{\frac {1}{u+\eta_t} \, A_j(t)\, \frac
{1}{u+\eta_t}\, A_j(t)\right\}
\\ {}
+ \rmd t \Tr\left\{B(t) \left( \ln \eta_{t}+ \int_0^{+\infty} \frac{\eta_t}{(u+\eta_t)^2}
\,\rmd u\right)\right\}.
\end{multline*}
By computing the integral we get
\begin{multline*}
\Tr\left\{B(t) \left( \ln \eta_{t}+ \int_0^{+\infty} \frac{\eta_t}{(u+\eta_t)^2}
\,\rmd u\right)\right\} =\Tr\left\{B(t) \left( \ln \eta_{t}+ \openone\right)\right\}
\\ {}=\Tr\left\{B(t) \ln \eta_{t}\right\}
= \sum_k\Tr \left\{ \left[\nu_k(t) - J_k(t)\right]\eta_t\ln \eta_t\right\} \\ {} +
\sum_j\Tr \left\{ \left[R_j(t)\eta_t R_j(t)^*- R_j(t)^*R_j(t)\eta_t\right]\ln
\eta_t\right\}
\\ {}+ \sum_l\Tr \left\{
\left[L_l(t)\eta_t L_l(t)^*- L_l(t)^*L_l(t)\eta_t\right]\ln \eta_t\right\}
\end{multline*}
and by using the integration by parts with $\frac 1 {(u+\eta_t)^2}= -\frac{\rmd\ } {\rmd
u}\,\frac 1 {u+\eta_t}$ we have also
\begin{multline*}
\sum_j \int_0^{+\infty}\rmd u \Tr\left\{\frac {1}{u+\eta_t} \, A_j(t)\, \frac {1}{u+\eta_t}\,
A_j(t)-\frac {\eta_t}{(u+\eta_t)^2} \, A_j(t)\, \frac {1}{u+\eta_t}\, A_j(t)\right\}
\\ {}=
\sum_j  \int_0^{+\infty}\rmd u \, \Tr \left\{ A_j(t) \, \frac 1 {(u+\eta_t)^2} \, A_j(t)\,
\frac u {u+\eta_t}\right\}
\\ {}=\sum_j  \int_0^{+\infty}\rmd u \, \Tr \left\{ A_j(t) \, \frac
{\eta_t} {(u+\eta_t)^2} \, A_j(t)\, \frac 1 {u+\eta_t}\right\}
\\ {}=
\frac 1 2 \sum_j \int_0^{+\infty}\rmd u \Tr\left\{\frac {1}{u+\eta_t} \, A_j(t)\, \frac
{1}{u+\eta_t}\, A_j(t)\right\}.
\end{multline*}

From the previous formulae we have the final expression
\begin{multline}\label{eq:res1}
- \frac{\rmd\ }{\rmd u}\,\E_{P_T}[S_\rmq(\rho^t_u)]\Big|_{u=t^+}=\sum_k\Tr \left\{
\mathcal{J}_k(t)[\eta_t]\ln \frac{ \mathcal{J}_k(t)[\eta_{t}]}{\nu_k(t)}  - J_k(t)\eta_t\ln
\eta_t\right\}
\\ {}+ \sum_j\Tr \left\{
R_j(t)\eta_t \big[R_j(t)^*,\,\ln \eta_t\big]\right\} + \sum_l\Tr \left\{ L_l(t)\eta_t\big[
L_l(t)^*,\, \ln \eta_t\big]\right\}
\\ {}+
\frac 1 2 \sum_j \bigg( \int_0^{+\infty}\rmd u\Tr \bigg\{R_j(t)^*\frac
{\eta_t}{u+\eta_t}R_j(t)^*\frac {\eta_t}{u+\eta_t} + \frac {\eta_t}{u+\eta_t}R_j(t)\frac
{\eta_t}{u+\eta_t}R_j(t) \\ {}+ \frac {2}{u+\eta_t}R_j(t)\frac {{\eta_t}^2}{u+\eta_t}R_j(t)^*
\bigg\}-n_j(t)^2\bigg).
\end{multline}

Analogously we get
\begin{multline}\label{eq:res2}
\frac{\rmd\ }{\rmd t}\,\E_{P_T}[S_\rmq(\rho_t)]=-\sum_k\E_{P_T}\left[\Tr \left\{
\mathcal{J}_k(t)[\rho_t]\ln \frac{ \mathcal{J}_k(t)[\rho_t]}{\mu_k(t)}  - J_k(t)\rho_t\ln
\rho_t\right\}\right]
\\ {}- \sum_j\E_{P_T}\left[\Tr \left\{
R_j(t)\rho_t \big[R_j(t)^*,\,\ln \rho_t\big]\right\}\right]- \sum_l\E_{P_T}\left[\Tr \left\{
L_l(t)\rho_t\big[ L_l(t)^*,\, \ln \rho_t\big]\right\}\right]
\\ {}-
\frac 1 2 \sum_j \E_{P_T}\bigg[ \int_0^{+\infty}\rmd u\Tr \bigg\{R_j(t)^*\frac
{\rho_t}{u+\rho_t}R_j(t)^*\frac {\rho_t}{u+\rho_t} \\ {}+ \frac {\rho_t}{u+\rho_t}R_j(t)\frac
{\rho_t}{u+\rho_t}R_j(t) + \frac {2}{u+\rho_t}R_j(t)\frac {{\rho_t}^2}{u+\rho_t}R_j(t)^*
\bigg\}-m_j(t)^2\bigg].
\end{multline}

By \eqref{eq:res1}, \eqref{eq:res2}, \eqref{eq:res3} we get the statement of the Proposition.
\end{proof}

\begin{cor}
A sufficient condition to have the equality in the main bound
\begin{equation}\label{eq:equality}
\frac{\rmd\ }{\rmd t}\, S_\rmc(p_t\| q_t)=  \frac{\rmd\ }{\rmd t}\,\E_{P_T}[S_\rmq(\rho_t)]-
\frac{\rmd\ }{\rmd u}\,\E_{P_T}[S_\rmq(\rho^t_u)]\Big|_{u=t^+}
\end{equation}
is to have $P_T$-a.s.\ in $\omega$ ($T\geq t$)
\begin{equation}\label{eq:commutation}
[V^r_k(t),\rho_t(\omega)]=0\,, \quad [R_j(t),\rho_t(\omega)]=0\,, \quad
[L_l(t),\rho_t(\omega)]=0\,, \quad \forall r,k,j,l.
\end{equation}

In the autonomous case, i.e.\ when $H, V^r_k, R_j, L_l$ are time independent, we have that the
conditions
\begin{equation}\label{eq:auto}
[H,\eta_0]=0, \quad [V^r_k,\eta_0]=0\,, \quad [R_j,\eta_0]=0\,, \quad [L_l,\eta_0]=0\,, \quad
\forall r,k,j,l,
\end{equation}
imply Eqs.\ \eqref{eq:commutation} and \eqref{eq:equality} $\forall t\geq 0$.
\end{cor}
\begin{proof}
By the commutation relations \eqref{eq:commutation} we get
\begin{multline*}
J_k(t)\rho_t\left(\ln \rho_t- \ln \eta_t\right) -\mathcal{J}_k(t)[\rho_t]\left(\ln
\mathcal{J}_k(t)[\rho_t] -\ln \mathcal{J}_k(t)[\eta_t]\right)\\ {}= J_k(t)\rho_t\left(\ln
\rho_t- \ln \eta_t-\ln J_k(t)\rho_t +\ln J_k(t)\eta_t\right)=0
\end{multline*}
and the first term in Eq.\ \eqref{eq:explicitb} vanishes.

By Eq.\ \eqref{eq:commutation} also the second and third term in Eq.\ \eqref{eq:explicitb}
vanish because they explicitly involve vanishing commutators.

Finally, let us consider one of the summands in \eqref{eq:explicitb}. We have
\begin{multline*}
\E_{P_T}\bigg[ \int_0^{+\infty}\rmd u\Tr \bigg\{R_j(t)^*\frac {\rho_t}{u+\rho_t}R_j(t)^*\frac
{\rho_t}{u+\rho_t}\bigg\}\bigg] \\ {}= \E_{P_T}\bigg[ \int_0^{+\infty}\rmd u\Tr
\bigg\{R_j(t)^{*2}\frac {{\rho_t}^2}{(u+\rho_t)^2}\bigg\}\bigg]
\\ {}=
\E_{P_T}\left[ \Tr \left\{R_j(t)^{*2}\rho_t\right\}\right] = \Tr
\left\{R_j(t)^{*2}\eta_t\right\}.
\end{multline*}
Similar formulae hold also for the other summands and the last term vanishes too.

In the autonomous case, when the initial state $\eta_0\equiv \rho_0$ commutes with all the
operators involved in the evolution equations, we get that Eqs.\ \eqref{eq:commutation} hold
for all $t$ and the conclusion follows from the first part of the corollary.
\end{proof}

\section*{Acknowledgments}
\noindent Work supported by the \emph{European Community's Human Potential Programme} under
contract HPRN-CT-2002-00279, QP-Applications.


\begin{thebibliography}{99}
\bibitem{BarLP83}
A.~Barchielli, L.~Lanz, G.~M.~Prosperi, {\sl Statistics of continuous trajectories in quantum
mechanics: Operation valued stochastic processes}, Found.\ Phys.\ {\bf 13} (1983) 779--812.
\bibitem{BarB91}
A.~Barchielli, V.~P.~Belavkin, \textsl{Measurements continuous in time and a posteriori states
in quantum mechanics}, J.\ Phys.\ A: Math.\ Gen. \textbf{24} (1991) 1495--1514.
\bibitem{Bar01}
A.~Barchielli,  \textsl{Entropy and information gain in quantum continual measurements}, in
P.\ Tombesi and O.\ Hirota (eds.), \textit{Quantum Communication, Computing, and Measurement
3} (Kluwer, New York, 2001) pp.\ 49--57; quant-ph/0012115.
\bibitem{BarL04}
A.~Barchielli, G.~Lupieri, \textsl{Instrumental processes, entropies, information in quantum
continual measurements}, in O.~Hirota (ed.), \textit{Quantum Information, Statistics,
Probability} (Rinton, Princeton, 2004) pp.\ 30--43; Quantum Inform.\ Compu.\ \textbf{4} (2004)
437--449; quant-ph/0401114.
\bibitem{BarLlevico}
A.~Barchielli, G.~Lupieri, \textsl{Entropic bounds and
continual measurements}, to appear in Quantum Probability Series QP-PQ, World Scientific;
quant-ph/0511090.
\bibitem{OhyP93}
M.\ Ohya, D.\ Petz, \textit{Quantum Entropy and Its Use} (Springer, Berlin, 1993).
\bibitem{Dix57}
J.\ Dixmier, \textit{Les Alg\`ebres d'Op\'erateurs dans l'Espace Hilbertien}
(Gauthier-Villars, Paris, 1957).
\bibitem{Dav76}
E.~B.~Davies, \textit{Quantum Theory of Open Systems} (Academic Press, London, 1976).
\bibitem{Oza84}
M.~Ozawa, \textsl{Quantum measuring processes of continuous observables}, J.\ Math.\ Phys.\
\textbf{25} (1984) 79--87.
\bibitem{BarL05QP}
A.~Barchielli, G.~Lupieri, \textsl{Instruments and mutual entropies in quantum information
theory}, to appear in Banach Center Publications; quant-ph/0412116.
\bibitem{Oza85a}
M.~Ozawa, \textsl{Conditional probability and a posteriori states in quantum mechanics},
Publ.\ R.I.M.S.\ Kyoto Univ.\ \textbf{21} (1985) 279--295.
\bibitem{Hol73}
A.~S.~Holevo, \textsl{Some estimates for the amount of information transmittable by a quantum
communication channel}, Probl.\ Inform.\ Transm.\ \textbf{9} no.~3  (1973) 177--183 (Engl.\
transl.: 1975).
\bibitem{HolS04}
A.~S.~Holevo, M.~E.~Shirokov, \textsl{Continuous ensembles and the $\chi$-capacity of
infinite-dimensional channels}, quant-ph/0408176 (2004).
\bibitem{BarL06qic}
A.~Barchielli, G.~Lupieri, \textsl{Quantum measurements and entropic bounds on information
transmission}, Quantum Information and Computation \textbf{6} (2006) 16--45; quant-ph/0505090.
\bibitem{BarHL93}
A.\ Barchielli, A.\ S.\ Holevo, G.\ Lupieri, {\sl An analogue of Hunt's representation theorem
in quantum probability}, J.\ Theor.\ Probab.\ {\bf 6} (1993) 231--265.
\bibitem{BarH95}
A.~Barchielli, A.~S.~Holevo, {\sl Constructing quantum measurement processes via classical
stochastic calculus}, Stoch. Proc. Appl. {\bf 58} (1995) 293--317.
\bibitem{CovT91}
T.\ M.\ Cover, J.\ A.\ Thomas, \textit{Elements of Information Theory} (Wiley, New York,
1991).

\end{thebibliography}
\end{document}